\documentclass[10pt,letterpaper]{article}

\usepackage[top=0.85in,left=2.75in,footskip=0.75in]{geometry}

\usepackage{changepage}

\usepackage[utf8]{inputenc}
\usepackage[brazilian,english]{babel}

\usepackage{textcomp,marvosym}

\usepackage{fixltx2e}

\usepackage{amsmath,amssymb}

\usepackage{cite}

\usepackage{nameref,hyperref}
\hypersetup{
        colorlinks,
        linkcolor={red!50!black},
        citecolor={blue!50!black},
        urlcolor={blue!80!black}
}

\usepackage{microtype}
\DisableLigatures[f]{encoding = *, family = * }

\usepackage{grffile}
\usepackage{etex}
\usepackage{morefloats}
\usepackage[section]{placeins}
\usepackage{rotating}
\usepackage{xcolor}       
\usepackage{listings}
\definecolor{dkgreen}{rgb}{0,0.6,0}
\definecolor{gray}{rgb}{0.5,0.5,0.5}
\definecolor{light-gray}{rgb}{0.97,0.97,0.97}

\lstdefinelanguage{owlms}
    {morekeywords={xsd,owl,xml,dc,rdf,skos,description,PlainLiteral,int,float,
        some,only,value,min,exactly,max,and,or,not,
        Prefix,Ontology,Import,Individual,Facts,Types,Class,
        DataProperty,ObjectProperty,AnnotationProperty,Annotations,
        DifferentIndividuals,SubClassOf,EquivalentTo,DisjointWith,DisjointUnionOf,SubPropertyOf,DisjointClasses,DisjointProperties,
        Symmetric,Asymmetric,Reflexive,Irreflexive,Transitive,Functional,InverseFunctional,
        Characteristics,Range,Domain,Datatype},
     basicstyle=\scriptsize\ttfamily,
     backgroundcolor=\color{light-gray},
     keywordstyle=\color{blue},
     commentstyle=\color{gray},
     stringstyle=\color{dkgreen},
     numbers=left,
     numberstyle=\tiny\color{gray},
     stepnumber=1,
     numbersep=10pt,
     tabsize=2,
     showspaces=false,
     showstringspaces=false,
     breaklines=true,                           
     sensitive=true,                            
     morecomment=[l][commentstyle]{\#},         
     morestring=[b]",                           
literate=%
  {ó}{{\'o}}1
  {ã}{{\~a}}1
  {Ã¢}{{\^a}}1
  {õ}{{\~o}}1
  {á}{{\'a}}1
  {Ãº}{{\'u}}1
  {í}{{\'i}}1
  {é}{{\'e}}1
  {ç}{{\c{C}}}1
  {Ã}{{\~O}}1
  {Ã}{{\^E}}1
  {Ã³}{{\'o}}1
  {Ã }{{\`a}}1
  {Ã}{{\^A}}1
  {Ã´}{{\^o}}1
  {Ãª}{{\^e}}1
  {Ã§}{{\c{c}}}1
}

\raggedright
\usepackage[aboveskip=1pt,labelfont=bf,labelsep=period,justification=raggedright,singlelinecheck=off]{caption}

\makeatletter
\renewcommand{\@biblabel}[1]{\quad#1.}
\makeatother

\date{}

\usepackage{lastpage,fancyhdr,graphicx}
\fancyheadoffset[L]{2.25in}
\fancyfootoffset[L]{2.25in}
\newcommand{\ops}{\textsc{ops}}
\newcommand{\opsi}{O\textsc{ps}}
\newcommand{\vcps}{\textsc{vcps}}
\newcommand{\owl}{\textsc{owl}}

\newcommand{\sparql}{\textsc{s}par\textsc{ql}}
\newcommand{\bfo}{\textsc{bfo}}

\newcommand{\foaf}{\textsc{foaf}}
\newcommand{\ict}{\textsc{ict}}
\newcommand{\html}{\textsc{html}}
\newcommand{\node}{\textsc{n}ode.js}
\newcommand{\facebook}{\textsc{f}acebook}
\newcommand{\twitter}{\textsc{t}witter}
\newcommand{\wwwc}{\textsc{w3c}}
\newcommand{\skos}{\textsc{skos}}
\newcommand{\etherpad}{\textsc{e}therpad}
\newcommand{\ogp}{\textsc{ogp}}

\newcommand{\uri}{\textsc{uri}}

\newcommand{\urll}{\textsc{url}}
\newcommand{\ngo}{\textsc{ngo}}
\newcommand{\http}{\textsc{http}}
\newcommand{\opa}{\textsc{op}a}
\newcommand{\ocd}{\textsc{ocd}}
\newcommand{\ontologiaa}{\textsc{o}ntologiaa}
\newcommand{\obs}{\textsc{obs}}
\newcommand{\pubby}{\textsc{p}ubby}
\newcommand{\rdf}{\textsc{rdf}}
\newcommand{\mysql}{\textsc{m}y\textsc{sql}}
\newcommand{\aan}{\textsc{aa}}
\newcommand{\cidadedemocratica}{\textsc{c}idade \textsc{d}emocr\'atica}
\newcommand{\participa}{\textsc{p}articipa.br}
\newcommand{\ontop}{\textsc{o}n\textsc{t}op}
\newcommand{\quest}{\textsc{q}uest}
\newcommand{\webprotege}{\textsc{w}ebprotege}
\newcommand{\obda}{\textsc{obda}}
\newcommand{\pnud}{\textsc{undp}}

\newcommand{\vbs}{\textsc{vbs}}
\newcommand{\lod}{\textsc{lod}}
\newcommand{\corais}{\textsc{c}orais}
\newcommand{\serpro}{\textsc{s}erpro}
\newcommand{\python}{\textsc{p}ython}
\newcommand{\protege}{\textsc{p}rot\`eg\`e}

\usepackage{xr}
\usepackage[section] {placeins}

\begin{document}
\vspace*{0.35in}

\begin{flushleft}
{\Large
\textbf\newline{Social Participation Ontology: community documentation, enhancements and use examples}
}
\newline
\\
Renato Fabbri\textsuperscript{1},
Henrique Parra Parra Filho\textsuperscript{2},
Rodrigo Bandeira de Luna\textsuperscript{2},
Ricardo Augusto Poppi Martins\textsuperscript{3},
Flor Karina Mamani Amanqui\textsuperscript{4},
Dilvan de Abreu Moreira\textsuperscript{4}
Osvaldo Novais de Oliveira Junior\textsuperscript{1},
\\
\bigskip
\bf{1} S\~ao Carlos Institute of Physics, University of S\~ao Paulo, CP 369, 13560-970, S\~ao Carlos, SP, Brazil, \url{fabbri@usp.br}
\\
\bf{2} Cidade Democr\'atica, Instituto Seva, S\~ao Paulo, SP, Brazil
\\
\bf{3} National Secretariat of Social Participation, General Secretariat of the Republic, Bras\'ilia, Federal District, Brazil
\\
\bf{4}  Institute of Mathematical and Computer Sciences, University of São Paulo, S\~ao Carlos, SP, Brazil
\\
\bigskip

%
%



%
%

\end{flushleft}
\section*{Abstract}
    Participatory democracy advances in virtually all governments and especially in South America
    which exhibits a mixed culture and social predisposition.
    This article presents the ``Social Participation Ontology'' 
    (\ops\ from the Brazilian name \emph{Ontologia de Participa\c{c}\~ao Social})
    implemented in compliance with the Web Ontology Language standard (\owl)
    for fostering social participation, specially in virtual platforms.
    The entities and links of \ops\ were defined based on an extensive collaboration of specialists.
    It is shown that \ops\ is instrumental for information retrieval from the contents of the portal,
    both in terms of the actors (at various levels) as well as mechanisms and activities.
    Significantly, \ops\ is linked to other \owl\ ontologies as an upper ontology
    and via \foaf\ and \bfo\ as higher upper ontologies,
    which yields sound organization and access of knowledge and data.
    In order to illustrate the usefulness of \ops, we present results on ontological expansion
    and integration with other ontologies and data.
    Ongoing work involves further adoption of \ops\ by the official
    Brazilian federal portal for social participation and \ngo s,
    and further linkage to other ontologies for social participation.
%

\section{Introduction}\label{sec:into}
Easy access to social media is reshaping citizen participation in government affairs~\cite{socMed}. 
Information and communication technologies (\ict s) have exhibited such an impact on the way individuals interact
that it is giving birth to new organizational methods in social movements.
These changes can be observed, for example, in the 2010 Arab Spring and the 2013 Brazilian protests.
These events gathered millions of people and, although recent,
have shown direct and strong impact in governments and new laws,
and the forecast is an intensification of the process~\cite{digRev1,digRev2,digRev3}.
Concomitantly, electronic government initiatives are flourishing,
favored mainly by the ubiquity of Internet technologies (e.g. \html\ 5, \node, open source browsers)
and by the need for renewal of representative democracy practices. 
These initiatives have taken place in various platforms, including usual social networks (e.g. \facebook, \twitter)
and dedicated instances created by both government and civil society parties~\cite{socMed,pita2010arquitetura,barros2010alem,knowledge}.
A natural challenge arises: how to link information produced into an unified knowledge base.
This is being addressed, at the technology level, by semantic web developments.
Endorsed by World Wide Web Consortium (\wwwc), current semantic web technologies include~\cite{Sem1}:
\begin{itemize}
    \item reasoning by means of ontological specifications;
    \item linking data from different sources (e.g. databases);
    \item organization of domain knowledge for coherent consideration.
\end{itemize}
\noindent Key among these technologies, ontologies are considered one of the pillars of the semantic web.
An ontology is usually defined as a formal specification of a shared conceptualization~\cite{gruber}.
They give meaning to data and are useful for datasets available on the web 
to make them automatically retrievable and linkable with other datasets.
The \wwwc\ created the Web Ontology Language (\owl) as a standard to represent ontologies in the web.
The second version of the language is called \owl\ 2 and offers greater expressive power~\cite{owl2}.

In this context, to describe and give meaning to social participation,
the ``Common Vocabulary of Social Participation''
(\vcps\ from the Brazilian name \emph{Vocabul\'ario Comum de Participa\c{c}\~ao Social})
was proposed as a joint effort of Latin America academic, civil and governmental groups~\cite{corais}.
Although started in 2012, a recent initiative for ontological developments,
it already yielded relevant material, including a public preliminary \owl\ ontology with a concise taxonomy.
Also important are the reference documents reporting results from a first working phase, from July to December, 2012. 
As stated by the community, \vcps\ was propelled by three goals:
1) to ease adoption of the vocabulary;
2) to stimulate the creation of public tools to understand, visualize and summarize how participation is happening;
3) to meet the need of participative initiatives to open and link their data. 
It is important to notice that \vcps,
an ontology, was called a \emph{vocabulary} both to ease understanding of the general public and because it started as a vocabulary.
The present article presents the ``Ontology of Social Participation'' 
(\ops\ from the Brazilian name \emph{Ontologia de Participa\c{c}\~ao Social}),
based on \vcps, in which the term \emph{vocabulary} was substituted by the term \emph{ontology} for the following reasons:
\begin{itemize}
    \item The usage of the word ``vocabulary'' can lead to confusion is some situations as \ops\ is an \owl\ ontology
    (not, for example, a \skos\ vocabulary).
    \item Documentation seems inconsistent when an ontology is repeatedly called a vocabulary.
    \item \opsi\ is, in fact, an ontology, with a vocabulary,
    a taxonomic organization and properties further relating the terms.
    \item This coherent naming is a prerequisite for academic acceptance and further formal adoption by government instances,
    such as the Brazilian Federal Portal of Open Data~\cite{dadosGov} and the Brazilian Federal Portal of Social Participation~\cite{participa}.
\end{itemize}

\noindent Also, the term \emph{common} was dropped when \ops\ was conceived,
as the term is redundant for an \owl\ ontology.
The \vcps\ presented other difficulties, such as missing classes,
incorrect \uri\ specifications (containing spaces), some logical flaws,
and unnecessary out-of-standards restrictions.
This were all solved within \ops\ (to the extent authors were able to, of course).

Next section presents \ops\ and modifications made from \vcps\ to \ops: class and property names and labels,
class restrictions and property axioms.
Section~\ref{ospUtil} is dedicated to \ops\ usage: dereferencing, \sparql\ queries,
a toy \ops\ expansion, discursive example of usage, and use cases from government,
civil society and academic parties.
Concluding remarks are stated with future work, in Section~\ref{conc}. 
The Appendixes hold directions for the scripts that are used to obtain \ops,
notes on restrictions that were in \vcps\ but withdrawn from \ops\, and an inspection of the first-hand
documentation about \vcps\ (reference textual documents, images, \owl\ code, blog posts, discussions and \etherpad s).

\section{OPS: the Ontology of Social Participation}\label{exp}
\begin{figure*}
    \centering
    \includegraphics[width=0.9\textwidth]{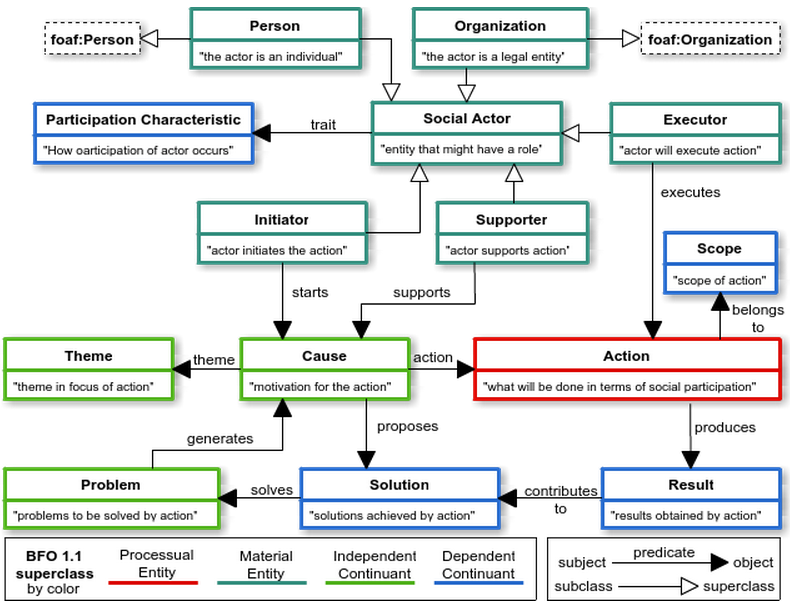}
    \caption{Diagram representation of the Ontology of Social Participation (\ops).
    Arrows with white heads indicate ``is a'' relations (subclass points to superclass).
    Arrows with black heads indicate property relations from subject to object .
    All property relations yield existential restrictions,
    with the exception of the ``has characteristic'' property,
    that does not yield restriction.
    Upper ontologies \bfo\ and \foaf\ are asserted through color (\bfo) and dashed boxes (\foaf).}
    \label{fig:v1}
\end{figure*}
This section makes considerations about \ops\ label standardization and implemented classes,
properties and restrictions.
Features present in Figure~\ref{fig:v1}, but not present in \vcps,
are fully described in Section~\ref{impl}.
Examples of usage are addressed in Section~\ref{ospUtil}. 

\subsection{Standardization and implemented features}\label{impl}
Without explicit criteria,
\vcps\ \uri\ was \url{http://lumii.lv/ontologies/Corais.owl}.
\opsi\ \uri\ was chosen to be \url{http://purl.org/socialparticipation/ops} for the following reasons:
\begin{itemize}
    \item This \uri\ is directly related to the ontology name (\ops).
    \item This \uri, also an \urll, is independent from government and other political associations.
    This is important to coalesce interested parties: the Brazilian Federal Social Participation Portal\cite{participa},
    the Brazilian government repository of vocabularies and ontologies~\cite{vocab},
    academic groups,  \ngo s, and non-organized civil society.
    \item Derived \uri s, when reached via \http, can be redirected to where current documentation is held,
    as it is hosted by \url{http://purl.org}.
\end{itemize}
Labels in the languages of interest should be written in label fields.
Even so, we advocate that \ops\ class names should be friendly to users,
bearing the attention not to take the class name as the label or as a meaning restriction.
For standardization, all classes are written in CamelCase~\cite{cc} in plain English to ease internationalization,
adoption and maturation.
Labels are written in English, Portuguese and Spanish.
Accordingly, class names changed from \vcps\ to \ops\,
received labels (\texttt{rdfs:label}) in each language
and a textual short explanation (\texttt{rdfs:comment}) in English. 
Table~\ref{ospClasses} exposes all classes is \ops.

\opsi\ property names fit headlessCamelCase~\cite{cc} format,
are readable in English (to ease internationalization, adoption and maturation),
and some of them have defined domains and ranges.
Table~\ref{ospProps} is dedicated to \ops\ properties, with labels in English, Portuguese and Spanish.

In the first versions of \ops, all properties yielded existential restrictions,
except {\tt ops:hasParticipationCharacteristic}.
Although such efforts were aimed at handling a better defined \ops,
further developments and discussions revealed that these restrictions made \ops\ rigid,
a bit more complicated, and was not of much help,
at least in this stage of \ops\ development and adoption.
The result is that all restrictions were removed.
Appendix~\ref{ap:restr} and Table~\ref{ospRestr} exposes all dropped restrictions from \vcps\ to \ops.

\begin{figure*}
    \centering
    \includegraphics[width=\textwidth]{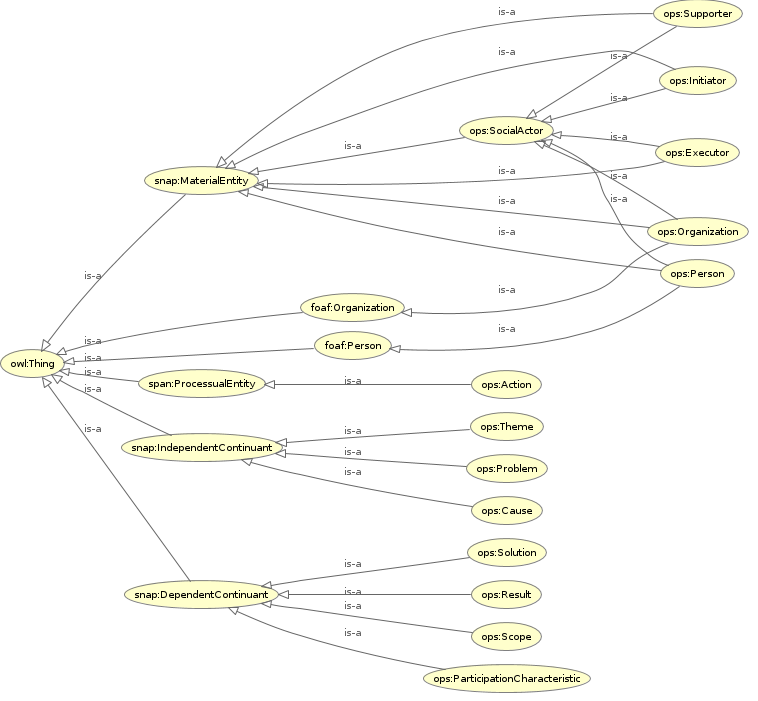}
    \caption{A taxonomic tree of the Ontology of Social Participation (\ops).
    This image was rendered inside \protege, with the \ops\ \owl\ code.
    Figure~\ref{fig:v1} is more informative, but this diagram is more standard and might be simpler for the newcomer.
    Note that the taxonomic tree does not present any information about properties further linking these classes.
    The prefixes {\tt snap:} and {\tt span:} are \bfo\ specific for {\tt snap}shot and {\tt span}ning-time.}
    \label{fig:owlCC}
\end{figure*}

An inspection of \vcps\ reveals a contradiction present in Figure~\ref{fig:diaorig}:
a role cannot execute, initiate or support a cause.
These are done by the social actor.
Therefore, to maintain \vcps\ directions and obtain a sound ontology,
{\tt ops:Role} was removed from \ops.
The result can be seen directly in Figures~\ref{fig:v1} and~\ref{fig:owlCC}.

\begin{table*}[!h]
\begin{adjustwidth}{-2.25in}{0in} 
    \footnotesize
  \centering
  \caption{Classes of the \ops\ (Ontology of Social Participation).
  These are core concepts in the ontology.
  Along with the taxonomic structure exposed in Figure~\ref{fig:owlCC},
  these classes are related by the properties in Table~\ref{ospProps}.
  In boldface is the \vcps\ class name ``{\tt vcps:Espa\c{c}o de A\c{c}\~ao}'',
  which caused original \vcps\ \owl\ to be pointed as corrupted by \protege\ (spaces are not allowed in \uri s).
  Also, class {\tt vcps:Role} was dropped as it yields logical problens in \vcps.}
  \begin{tabular}{|p{1.8cm}|p{1.6cm}||p{2.2cm}|p{2.2cm}|p{1.8cm}||p{4cm}||p{3cm}|}\hline
      {\bf OPS class name} & {\bf VCPS class name} & {\bf pt-br label} & {\bf es label} & {\bf en label} & {\bf definition} & {\bf upper ontology classes} \\\hline\hline
      Person & Pessoa & Pessoa & Persona & Person & a person (social actor is a person) & {\tt bfo:'Material Entity'}, {\tt foaf:Person} \\ \hline
      Organization & Organizacao & Organiza\c{c}\~ao & Organizaci\'on & Organization & social actor is a group of individuals, organized formally or informally (e.g. movements, collectives) & {\tt bfo:'Material Entity'}, {\tt foaf:Organization} \\ \hline \hline
Executor & Executor & Executor & Ejecutor & Executor & performs action directly and is responsible for results & {\tt bfo:'Material Entity'} \\ \hline
Initiator & Iniciador & Iniciador & Iniciador & Initiator & originates cause, individually or collaborativelly & {\tt bfo:'Material Entity'} \\ \hline
      Supporter & Apoiador & Apoiador & Apoyador & Supporter & supports cause with resources of any kind (e.g. cognitive, financial, equipments) & {\tt bfo:'Material Entity'} \\ \hline
SocialActor & Ator & Ator Social & Actor Social & Social Actor & entity that might have a participatory role & {\tt bfo:'Material Entity'} \\ \hline \hline
Participation-Characteristic & NivelDe-Participacao & Caracter\'istica de Participa\c{c}\~ao & Caracter\'istica de Participaci\'on & Participation Characteristic & the way the participation of the specific actor is happening & {\tt bfo:'Dependent Continuant'} \\\hline
Cause & Causa & Causa & Causa &  Cause & the motivation for Action & {\tt bfo:'Dependent Continuant'} \\\hline
      Scope &{\bf Espa\c{c}o de A\c{c}\~ao} & Escopo & Ambito & Scope & the scope os Action & {\tt bfo:'Dependent Continuant'} \\\hline
Result & Resultados & Resultado & Resultado & Result & the result obtained with action & {\tt bfo:'Dependent Continuant'} \\\hline
      Solution & Solucao & Solu\c{c}\~ao & Soluci\'on & Solution & solution achieved with Action & {\tt bfo:'Dependent Continuant'} \\\hline \hline
      Problem & Problema & Problema & Problema & Problem & the problem that the Action aims to solve & {\tt bfo:'Independent Continuant'} \\\hline
Theme & Tema & Tema & Tema & Theme & the theme in focus by Action & {\tt bfo:'Independent Continuant'} \\\hline \hline
Action & Acao & A\c{c}\~ao & Acci\'on & Action & what is done in terms os social participation & {\tt bfo:'Processual Entity'} \\\hline\hline
\textcolor{red}{dropped class} & Papel & -//-  & -//- & -//- & the role of the actor & -//- \\ \hline
  \end{tabular}
  \label{ospClasses}
\end{adjustwidth}
\end{table*}

\begin{table*}[!h]
\begin{adjustwidth}{-2.25in}{0in} 
  \caption{Properties of the \ops\ (Ontology of Social Participation) along original \vcps\ names.
  Only a few ranges were established, and no domain, as these were not regarded as useful at time and,
  without them, \ops\ can be used more freely.}
  \begin{tabular}{|l|l||p{2.2cm}|p{2.2cm}|p{1.8cm}||l|p{1.2cm}|}\hline
{\bf OSP property name} & {\bf VCPS property name} & {\bf pt-br label} & {\bf es label} & {\bf en label} & {\bf domain} & {\bf range} \\\hline\hline
theme & possuiTemaAssociado & tema & tema & theme & -//- & Theme \\ \hline
belongsTo & pertenceAoEspaco & pertence ao & pertence al & belongs to & -//- & Scope \\ \hline
action & possuiAcao & a\c{c}\~ao & acci\'on & action & -//- & Action \\\hline\hline
supports & apoiaCausa & apoia & apoya & supports & -//- & -//- \\ \hline
contributesTo & compoeSolucao & contribui para & contribuye para & contributes to & -//- & -//- \\ \hline
executes & executaAcao & executa & ejecuta & executes & -//- & -//- \\ \hline
generates & geraCausa & gera & genera & generates & -//- & -//- \\ \hline
starts & iniciaCausa & inicia & inicializa & starts & -//- & -//- \\ \hline
solves & soluciona & soluciona & resuelve & solves & -//- & -//- \\ \hline
produces & produzResultado & produz & produce & produces & -//- & -//- \\\hline
proposes & propoeSolucao & prop\~oe & propone & proposes & -//- & -//- \\\hline
      trait & temNivelDeParticipacao & tra\c{c}o & rasgo & trait & -//- & -//- \\\hline\hline
{\color{red} dropped} & temPapel & -//-  & -//-  & -//-  & -//-  & -//- \\ \hline
  \end{tabular}
  \label{ospProps}
\end{adjustwidth}
\end{table*}

A comparison of the \vcps\ \owl\ code~\cite{owlCCPtg} 
with the diagram in Figure~\ref{fig:diaorig}, which reflects official \vcps\ documentation,
revealed that a class, two properties and three restrictions were not implemented in \vcps.
These were fully implemented in \ops\ before all restrictions were removed.
These are the missing class and properties 
(restrictions missing in \vcps\ and implemented in preliminary \ops\ versions are exposed in Appendix~\ref{ap:restr}):
\begin{itemize}
    \item Class: {\tt    ops:ParticipationCharacteristic}.
    \item Property: {\tt ops:hasRole}.
    \item Property: {\tt ops:composesSolution}.
\end{itemize}

\opsi\ is available online~\cite{owlOSP}.
To ease navigation of the ontology by interested parties,
it is also available in the Webprotege interface~\cite{owlOSPwp}.
The diagram of \ops ' taxonomic structure is exposed in Figure~\ref{fig:owlCC}.

Upper ontologies usage with \ops\ is under development and should receive a dedicated article,
as possibilities should be inspected carefully.
Pertinent and already used as upper ontologies for \ops\ are \foaf~\cite{foaf}
(for linking and describing people and things they do) and \bfo~\cite{bfo}
(``designed for use in supporting information retrieval, analysis and integration in scientific and other domains''
as stated on their documentation).
Properties were not related to upper ontologies as reasonable relations are still being searched for.
Upper ontologies classes related to each \ops\ class are also exposed in Table~\ref{ospClasses}.

Figure~\ref{fig:v1} is a complete diagram of current \ops:
classes, properties and relations to \foaf\ and \bfo.
Actually, Figure~\ref{fig:v1} is more informative than the \ops\ \owl\ code,
as restrictions were removed and not all properties have defined domain and range.
Therefore, the diagram is an important source of relations envisioned by \ops\ creators.

\section{OPS usefulness}\label{ospUtil}
\opsi\ is meant to be useful in at least two ways.
First, as a systematization of what is social participation to Latin America groups,
as conceived by \vcps.
Second, as a mean to ease linked data,
and enable integration of various instances for social participation.
An indicative of this pertinence is \opa, \ocd, \ontologiaa, and \obs,
ontologies that already uses \ops\ as upper ontology~\cite{pnud5}.

\opsi\ usage might not be obvious at first.
How is data linked? How is field knowledge organized? Why and by whom?
Core principles of \ops\ utility can be understood by the following observations:
\begin{itemize}
    \item Different participation instances have social actors,
    actions being developed, organizations involved, problems being tackled, etc.
    These can yield one consistent database by means of \ops\ usage.
    \item One can understand and share the mutually exclusive nature of being
    a paid or a voluntary contributor by observing the expanded version of \ops\ (see Section~\ref{downwards}).
    Also, noticing the fact that a mob can be very big or not,
    and that it can be convoked or not by a network,
    can make the field conceptualization more clear for a newcomer
    or ease discussions and problematization for senior researchers or politicians.
    \item Other fields of human knowledge and practice also have agents, problems and so forth.
    These can be linked to participatory data and ontologies by means of \ops.
    \item \opsi\ carry a conceptualization that is the product of contribution of many specialists.
    In other words, put aside the linked data aspects,
    it is useful as an organization of the knowledge related to the social participation field.
\end{itemize}

The rest of this section explores different \ops\ uses:
dereferencing, \sparql\ queries, expansion, fictional use cases that reflect its potential,
and real use cases.

\subsection{Usage of OPS through dereferencing}
\begin{figure}[!h]
    \centering
    \includegraphics[width=\columnwidth]{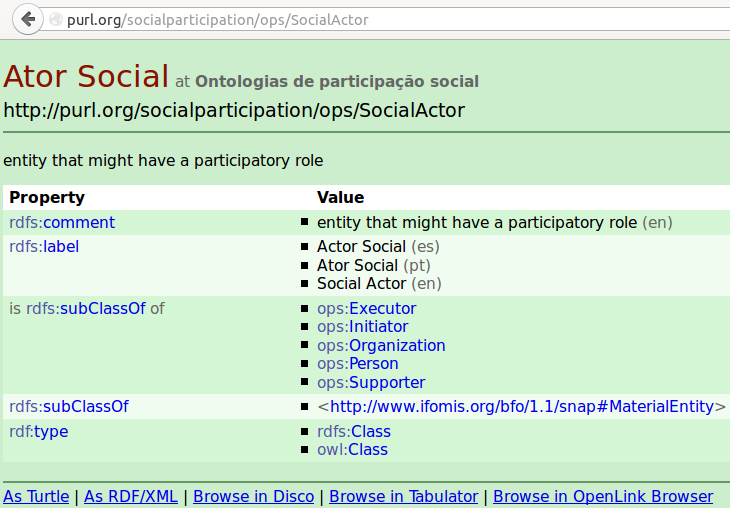}
    \caption{Dereferencing an \ops\ class:
    the \uri\ is also an \urll, which, reached by \http,
    returns information for the user as shown.
    Also, if the client is not an \html\ browser, but a crawler or another type of linked data application,
    \pubby\ delivers plain \rdf, not an user-friendly \html.}
    \label{fig:deref}
\end{figure}

All \ops\ classes and properties \uri s are accessible via \http.
A \pubby~\cite{pubby} instance delivers information like name,
labels and relations to other classes and properties.
As an example, the \uri\ \url{http://purl.org/socialparticipation/ops/SocialActor} returns information about this class,
as shown in Figure~\ref{fig:deref}.

\subsection{Usage of OPS through a SparQL endpoint}
\begin{figure}[!h]
    \centering
        \includegraphics[width=0.9\columnwidth]{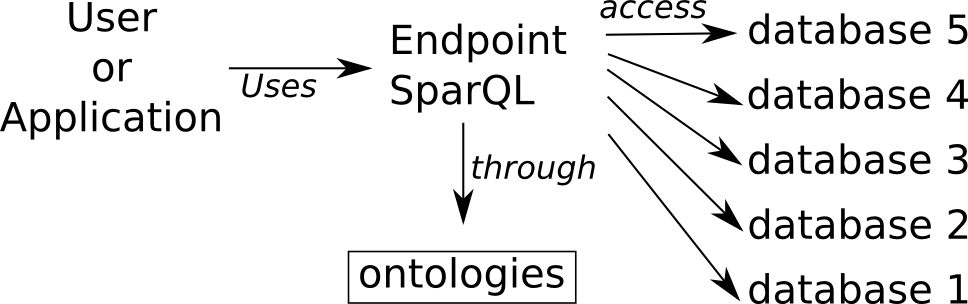}
    \caption{Scheme of the common use of ontologies for multiple databases integration.
    A user or application reaches a \sparql\ endpoint.
    This endpoint, through ontologies, delivers data from one or more databases.
    Nowadays, the most usual is to find only one database available at an endpoint,
    and this database is usually duplicated and not synchronized,
    but available as a (converted) triplestore.
    Even so,
    it is possible to access multiple ontologies and it is desirable that the databases have synchronous access,
    i.e. without need to convert data to \rdf\ triples beforehand.}
    \label{endpoint}
\end{figure}

Linking multiple databases is an \owl\ technology core purpose.
The standard way to access these data via ontologies is by using a \sparql\ endpoint.
Such endpoint delivers data from a triplestore (a collection of \rdf\ triples) or,
with more experimental technology, from relational database systems, such as a \mysql\ server 
(e.g. via \ontop/\quest~\cite{onTop}).
Either way, the query is the same:
the user or machine reaching the endpoint uses the \sparql\
protocol in order to retrieve information through semantic criteria.
Figure~\ref{endpoint} is a schematic representation of \obda\ (Ontology Based Database Access),
which is a common name for this multiple database access through ontologies.

Some examples of this usage can be given by \sparql\ queries and concise explanations:
\begin{itemize}
    \item \texttt{"select ?s where \{ ?s a ops:SocialActor \}"} or
    \texttt{"select ?s2 where \{ ?s2 a ops:Person \}"} or
    \texttt{"select ?s3 where \{ ?s3 a ops:Organization\}"}:
    the first query retrieves all social actors (returned in variable \texttt{?s}),
    be each a person, an organization, or something else;
    the second query retrieves all persons (variable \texttt{?s2});
    the third query retrieves all organizations (\texttt{?s3}).
    In a similar manner, one can retrieve all roles played, all executers, all initiators and all supporters.
    \item \texttt{"select ?s ?o where \{?s ops:starts ?o\}"}:
    this query retrieves all causes ({\tt ?o}) and their initiators ({\tt ?s})
    or whatever use of the {\tt ops:starts} property is taking place.
    \item \texttt{"select ?s ?s2 ?o ?o2 where \{?s a ops:Action . ?s ops:belongsTo ?o . ?s2 ops:executes ?s . ?s ops:produces ?o2\}"}:
    this query retrieves all Actions (\texttt{?s}) along their Action Fields (\texttt{?o}),
    their Executers (\texttt{?s2}) and their Results (\texttt{?o2}).
\end{itemize}

Noteworthy is that while {\tt opa:Participant} can be used to retrieve all \participa\ participants,
{\tt ocd:Participant} can be used to retrieve all \cidadedemocratica\ participants,
and {\tt aa:User} can be used to retrieve all \aan\ participants;
their upper ontology class {\tt ops:SocialActor} retrieves all of them
and relates these entities directly to the class of generic actors of social participation processes~\cite{pnud5}.

\subsection{OPS expansion}\label{downwards}
\begin{figure}[!h]
        \includegraphics[width=1.05\textwidth]{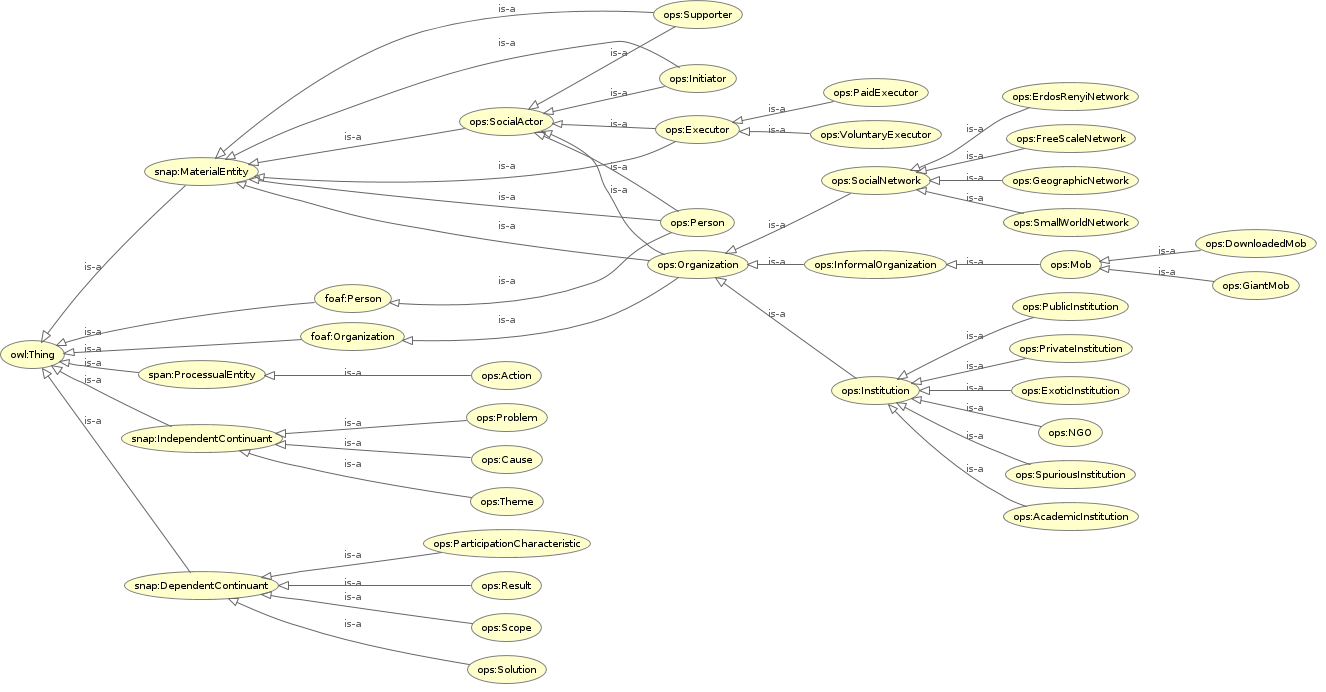}
        \caption{A taxonomic diagram of an example expanded instance of \ops.
	{\tt ops:PaidExecutor} is defined as being the subject of a
	\texttt{ops:receivesFrom} relation with a \texttt{ops:SocialActor};
	\texttt{ops:DownloadedMob} is defined by being the subject of a
	\texttt{ops:convoquedBy} relation with a \texttt{ops:SocialNetwork}.
	The \owl\ code is online for live editing~\cite{owlExp}.
	New classes added to this expanded version of \ops\ are in Table~\ref{ospFooClass}}
    \label{fig:owlExp}
    \end{figure}

\opsi\ matches \vcps\ online documentation~\cite{corais} as strictly as possible while maintaining logic coherence
and most useful relations.
In this context, to examplify the usefulness of the conceptualization
and possible additional classes,
an expanded \ops\ ontology is presented in this section 
and is uploaded to \webprotege~\cite{owlExp}.
Table~\ref{ospFooClass} is dedicated to these additional classes
while Figure~\ref{fig:owlExp} exposes the resulting taxonomic structure.

The property {\tt ops:receivesFrom} was added and has an inverse:
\texttt{ops:SocialActor ops:paysTo ops:Executor}.
Also, the {\tt ops:DownloadedMob} class is a defined class by the existential restriction:
\texttt{ops:Mob ops:convoquedBy ops:Network}, with a newly defined property \texttt{ops:convoquedBy}.

This is one of the numerous ways by which \ops\ might cope with further classes,
properties and restrictions.
This particular expansion was chosen as an example by direct observance of
\vcps\ documentation and recent social affairs,
such as the Brazilian protests.

\begin{table*}[!h]
\begin{adjustwidth}{-2.25in}{0in} 
  \centering
  \caption{New classes considered for an example expansion of the \ops.
  The taxonomic organization of these classes within \ops\ can be observed in Figure~\ref{fig:owlExp}.
  Further information is in Section~\ref{downwards}.}
  \begin{tabular}{|l|l|p{5.4cm}|p{5.3cm}|}\hline
{\bf new class} & {\bf subclass of} & {\bf description} & {\bf further notes} \\\hline\hline
SocialNetwork & Organization & a social structure made up of social actors (such as individuals or organizations) and a set of dyadic ties between these actors & --//--\\\hline
ScaleFreeNetwork & SocialNetwork & a network whose connectivity follows a power law distribution & disjoint with ErdosRenyiNetwork and GeographicNetwork. Usual social network structure \\ \hline
ErdosRenyiNetwork & SocialNetwork & also known as Binomial network, this network sets, with equal propability, an edge between each pair of nodes & disjoint with ScaleFreeNetwork and GeographicNetwork. Unusual social network structure \\\hline
GeographicNetwork & SocialNetwork & a network whose connectivity is related to the distance of nodes in a metric space & disjoint with both ScaleFreeNetwork and ErdosRenyiNetwork. Unusual social network structure \\\hline
SmallWorldNetwork & SocialNetwork & a network where most nodes can be reached from other nodes with few hops or steps & disjoint with GeographicNetwork. Usual social network structure \\\hline\hline
InformalOrganization & Organization & an organization that is not formalized & disjoint with Intitution \\ \hline
Mob & InformalOrganization & a crowd of individuals & --//-- \\\hline
GiantMob & Mob & a crowd with more than 10,000 individuals & --//--\\\hline
DownloadedMob & Mob & a Mob convoqued by a network & this is a defined class, by being a network and being the subject of the relation convoquedBy with object Network \\ \hline\hline
Institution & Organization & a mechanism of social order that governs a set of individuals & disjoint with InformalOrganization \\\hline
PublicInstitution & Institution & an institution backed through public funds and controlled by the state & disjoint with PrivateInstitution \\ \hline
PrivateInstitution & Institution & an institution backed through private fundings and controlled by private parties & disjoint with PublicInstitution \\ \hline
AcademicInstitution & Institution & an institution dedicated to education and research, which grants academic degrees & --//-- \\ \hline
NGO & Institution & a legally consituted corporation created by natural or legal people that operate independently from any form of government & --//-- \\\hline
SpuriousInstitution & Institution & an institution that holds prominent illegitimate or corrupt characteristcs & --//-- \\\hline
ExoticInstitution & Institution & an institution that does not fit previous classes or is characterized by very unique traces & --//-- \\\hline\hline
VoluntaryExecutor & Executor & an executor that receives no formal reward for the tasks & disjoint with PaidExecutor \\ \hline
PaidExecutor & Executor & an Executor that receives formal reward for the tasks accomplished & a defined class, being an Executor and the subject of a receivesFrom predicate with SocialActor as object\\ \hline
  \end{tabular}
  \label{ospFooClass}
\end{adjustwidth}
\end{table*}

\subsection{Fictional examples of usage}\label{examplesUsage}
The following fictional examples 
have the purpose of easing one to grasp why and how \ops\ can be used:
\begin{itemize}
    \item Suppose that a public \sparql\ endpoint unifies several participation instances by means of \ops\ 
    (we will see in Section~\ref{sec:real} that this is not really fictional).
    Thus, the total number of participants is publicly available ({\tt |ops:SocialActor|}).
    Also, depending on the platforms involved,
    one can observe how many of these participants are individuals ({\tt ops:Person}), 
    how many are organizations ({\tt ops:Organization}),
    and understand to which extent the corporative influence is explicit.
    One can observe how many of the participants are the same in each platform,
    and what roles they take,
    and make assumptions about how much the society is really participating
    or if these processes are manipulated by few agents ({\tt ops:SocialActor}).
    One can also gaze upon the problems being discussed and which solutions are being proposed,
    therefore easing the sense of what is being considered important and valid as public discussions.
    This list of possibilities is endless, specially when \ops\ variations and expansions are considered.
    \item Suppose a person, has a new proposal for a participatory system that uses \ops.
    She can acquire a concise understanding of the conceptualization involved
    and make very objective observations and deliver clear suggestions
    that relate directly to the systems used or envisioned.
    She can make an \ops\ variation or another ontology, as a way to confront paradigms.
    \item Suppose there is a system for exhibiting indicators about social participation 
    (e.g. how effective it has been, how wide is the scope of interests).
    Instances that are integrated by \ops\ can be queried for information and,
    for example, this system registers any organization involved as a social actor.
    Also, reflecting the expanded \ops\ exposed above, the system can register any mob involved,
    whose incidence was recorded in the database as a \texttt{DownloadedMob},
    as related to some social network (the network might be unknown).
\end{itemize}

\subsection{Real use cases}\label{sec:real}
\opsi\ is a recent ontology.
Even so, some real use cases can be pointed, from which are most notable:
\begin{itemize}
    \item The \pnud\ consultant contract 2013/00056, project BRA/12/018,
    was profoundly influenced by \ops~\cite{pnud5}.
    Within written products are other participatory ontologies,
    such as \opa, \ocd, \ontologiaa\ and \obs,
    which relates directly to \ops.
    Also, some methods for analyzing \ops\ related data and for resource recommendation were delivered.
    These developments were done by the computational physics researcher R. Fabbri (first author of the present article),
    in collaboration with other parties, specially the Brazilian General Secretariat of the Republic,
    University of Bras\'ilia researchers, and free software parties.
    Data from three participatory instances were triplified:
    \participa, \aan\ and \cidadedemocratica; all related to \ops.
    These linked data resources are available in \sparql\ endpoints and can be dereferenced,
    in a similar fashion as done for \ops.
    As these are all executed in research facilities,
    they might lack maintenance and should be kept by a dedicated team.
    \item Another \pnud\ contract was responsible for some advances in information technologies.
    A special case is a dedicated \ops\ expansion\cite{paulo6}.
    \item The Brazilian Federal Data Processing Service (\serpro) is triplifying \participa\ data 
    (the Brazilian federal social participation portal). Is this process, they as making direct use of \ops~\cite{tripSerpro}.
    \item The linked data Brazilian community had some maturing related to \ops\ beyond the ontologies developed,
    \pnud\ development and documentation, \serpro\ data triplification, and \vcps.
    Interested government, civil society and academic parties circulated \ops\
    documentation for conceptual and technological goals~\cite{circulaOps1,circulaOps2}.
\end{itemize}

\section{Concluding remarks and further work}\label{conc}
\opsi, based in \vcps,
yields initial steps in achieving an effective social participation ontology.
Community has registered activities and delivered reference documents,
including this present article which contains 
an organization of the community foundational conceptualizations,
\ops\ diligent development, uses,
upper ontologies and expansion of the ontological structures.

On the practical side, the use of this ontology or related developments
for the Brazilian federal participation portal (\participa~\cite{participa}) is a desirable reality,
as it implies usage and good maintenance.
Moreover, an ontology was done for \participa, based in the \ops: the \opa~\cite{pnud5} (Ontology of \participa).
This is confluent with the presidential Decree 8.243 that establishes a policy and commitment for social participation\cite{decree}.
In this context, presidential, ministerial and academic parties started formalizing current
legal participatory mechanisms (e.g. conferences, councils, forums, public consultations, round tables)
in ontological terms, which resulted in the Social Library Ontology 
(\obs\ from the Brazilian name Ontologia da Biblioteca Social)
and the Social Library Vocabulary (\vbs\ from the Brazilian name Vocabul\'ario da Biblioteca Social).
Hosting ontologies on Webprotege~\cite{webprotege} have become central,
as a way to share specific ontologies in a friendly environment and to collect feedback.

Further work involves observing community manifestations about \ops\ and this article,
accomplishing use by means of formal instances and civil society,
and studying upper ontologies for \ops.
The use of \ops\ (or a variant) in different instances
is being tackled for the creation of social participation 
indicators and easing participation processes.
Academic texts dedicated to the new participation ontologies 
(\ocd, \obs, \ontologiaa, \opa) which uses \ops\ as an upper ontology
should be written and submitted to peer review for enhancements and quality assurance~\cite{pnud5}.
Data related to these ontologies were found in relational databases 
and preliminary scripts were written to make them available as \rdf~\cite{datahub}.
A sound linkage of this data and the consequent incorporation
to the Linked Open Data (\lod~\cite{lod}) cloud is planed for a near future.
This should make Brazilian participative structures and data more relevant to the Giant Global Graph~\cite{ggg}.

\subsubsection*{Acknowledgments}
This article is deeply influenced by \protege\ and \bfo\ documentation.
Authors thank community and researchers related to these projects~\cite{protege,bfo}.
Authors thank \corais\ platform maintainers for their efforts 
in delivering a collaborative platform which gave birth to \vcps\ documentation.
Renato Fabbri is grateful to \textsc{cnp}q (process 140860/2013-4, project 870336/1997-5),
\pnud\ (contract 2013/00056, project BRA/12/018), \textsc{snas/sg-pr},
and the postgraduate committee of the \textsc{ifsc/usp}.

\appendix

\section{Script for obtaining current OPS}\label{ap:script}
Preliminary \ops\ was done with \protege\ software~\cite{protege}.
Current \ops\ is the output of a Python script~\cite{opsScript}.
Actually, this article, latex files, the \owl\ code, the \python\ script 
and auxiliary files are in a public git repository~\cite{opsRepo},
as is the common practice of the first author of this article.

\section{Restrictions in VCPS and initial OPS which were removed from current OPS}\label{ap:restr}
These restrictions were part of \vcps\ documentation but were not implemented in \vcps\ \owl\ code:
\begin{itemize}
    \item Restriction: {\tt ops:Role ops:hasParticipationCharacteristic some ops:ParticipationCharacteristic}.
    \item Restriction: {\tt ops:Results ops:composesSolution some ops:Solution}.
    \item Restriction: {\tt ops:Problem ops:generatesCause some ops:Cause}.
\end{itemize}

Present in preliminary \ops\ implementation,
all restrictions were removed from current \ops\ as explained in Section~\ref{impl}.
\begin{table}[!h]
  \centering
  \caption{Restrictions of the preliminary \ops: all restrictions are existential (\texttt{owl:someValuesFrom)}.}
  \begin{tabular}{|l|l|l|}\hline
{\bf subject} & {\bf predicate} & {\bf object} \\\hline\hline
Initiator    & starts        & Cause\\\hline
Supporter    & supports      & Cause\\\hline
Executer     & executes     & Action\\\hline
Solution     & solves      & Problem\\\hline
SocialActor  & hasRole            & Role\\\hline
Action       & produces    & Results\\\hline
Result       & contributesTo   & Solution\\\hline
Cause        & action          & Action\\\hline
Action       & belongsTo     & Scope\\\hline
Cause        & theme           & Theme\\\hline
Cause        & proposes   & Solution\\\hline
Problem      & generates     & Cause\\\hline
  \end{tabular}
  \label{ospRestr}
\end{table}

Such \owl\ restrictions are valid, to the best of chances,
for the final state of a participatory process 
(for example, in an arbitrary snapshot, a \texttt{SocialActor} may be not tied to a role).
This might lead to NULL or ``not yet defined'' field supplies.
Also, in \vcps, existential restrictions were written as ``min 1''. 
These were changed to the standard ``some'' existential restriction in preliminary \ops\
and were completely removed afterwards, as a way to avoid making \ops\ usage rigid and unnecessarily complicated.

%

\section{VCPS original documentation}\label{sec:orig}
From April to December, 2012, \vcps\ was first conceived.
In the online process, as registered by \corais\ platform, 66 users interacted,
6 of them were the most active~\cite{metodologia}.
Various materials were produced both as activity traces and as reference media.
This section is dedicated to these materials.

\subsection{Reference textual documents}\label{refDocs}
The main documents are:
\begin{itemize}
    \item ``Commented methodology''~\cite{metodologia}:
    this document describes the public process of \vcps\ conception.
    It is composed by brief inspections of forum topics,
    pointing both pertinent characteristics of the online collective process and ontological observations 
    (about classes and properties).
    Considerations are made about tightening relations with the Open Government Partnership (\ogp),
    an international initiative to foster transparency and open practices in governments worldwide~\cite{OGP},
    and the Brazilian formal action plan, as means to achieve ontology usage. 
    There is also a proposal of a systematic study of electronic government initiatives,
    so that the \vcps\ might be better contextualized.
    This document ends by proposing an agenda of meetings with academics, entrepreneurs and government parties.
    \item ``Conceptual modeling, version 0.1 (in natural language)''~\cite{conceptualMod}: 
    this document is a description, in English, of the \vcps.
    The introduction is mainly a collage of the document above~\cite{metodologia}.
    Both the itemized description of the ontology, and the considerations for its usage,
    are of great value as references.
    Figure~\ref{fig:v1} is heavily influenced by a diagram related to this document and further described in Section~\ref{sec:im}.
\end{itemize}

\subsection{Images}\label{sec:im}
\begin{figure}[!h]
    \centering
    \includegraphics[width=0.4\textwidth]{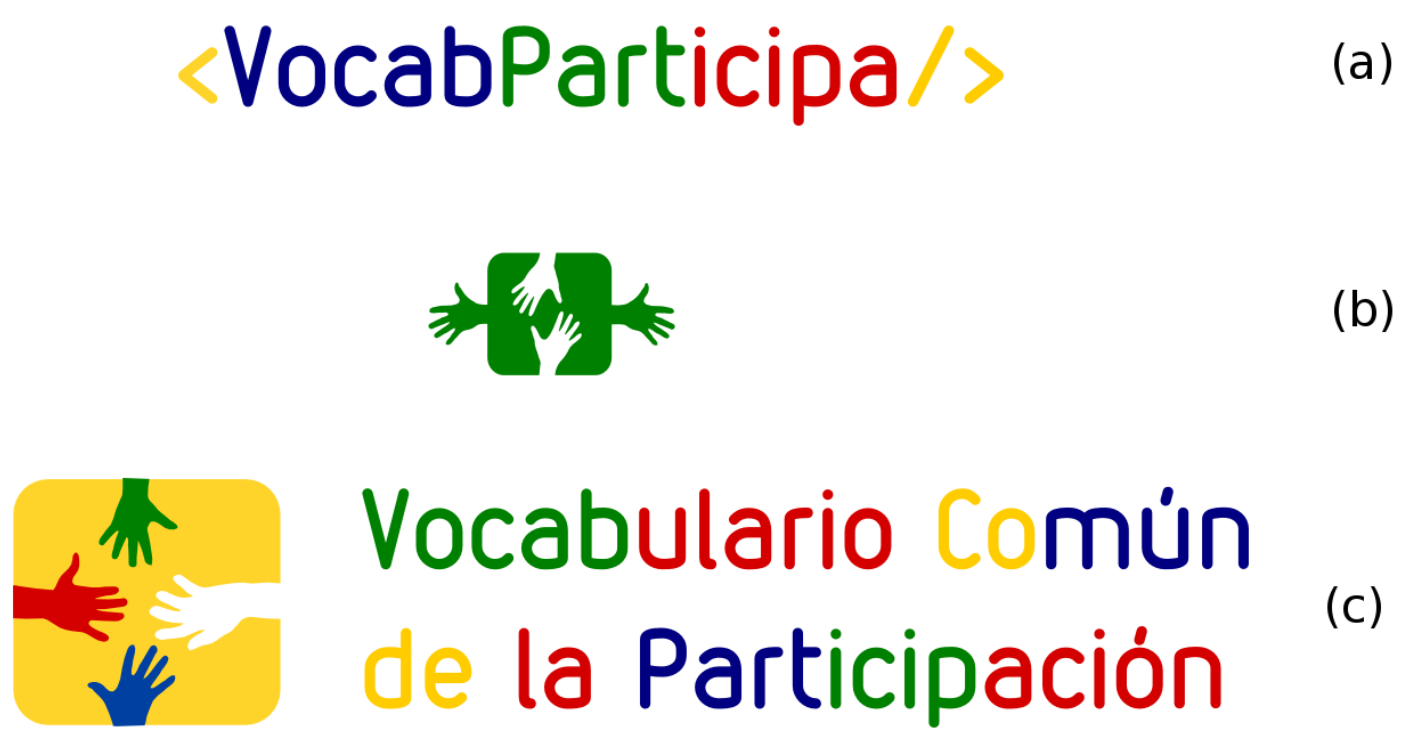}
    \caption{Some of the various logos for of the \vcps.
    (a) is a colored text logo proposal; (b) is a figurative logo; (c) is mixture of both ideas.
    It can be seen that these logos were conceived for the ontology when it was called a \emph{vocabulary}.
    Community documents reflect this nomenclature, which changed with first versions of this article,
    in the final months of 2013, and subsequent work.}
    \label{logo}
\end{figure}

There are various images associated with the ontology\footnote{Corais platform page with many images:
\url{http://corais.org/vocabulariodaparticipacao/node/1517}.}, most notably:
\begin{itemize}
    \item Various proposals for the \vcps\ logo, some of which are in Figure~\ref{logo}.
    \item Figure~\ref{fig:v1} shows an English version 
    of the original diagram of \vcps\ in the document~\cite{conceptualMod}.
    \item A diagram for general public consultations.
\end{itemize}

\subsection{OWL code of VCPS}\label{owl}
The \owl\ code of \vcps\ is online~\cite{owlVcps} and deprecated by \ops\ advent. 
The \vcps\ \owl\ code did not contain all relation from Figure~\ref{fig:v1}.
This is directly addressed in Section~\ref{exp},
which exposes the implementation of all relations in the \ops,
including \vcps\ \owl\ corrections and adjustments to best practices.
The complete and correct \ops\ is further contextualized in Section~\ref{ospUtil}.

\subsection{Blog Posts}
The \vcps\ blog
aggregates both important discussions and documents in no more than twenty posts to date.
All \owl\ code, final documents, public consultations, mental map and images are posted in the blog~\cite{coraisBlog}. 
The first post is from July 24, 2012.
Last post about \vcps\ is from May 7, 2013.
Most blog posts are from the first day (almost half of them).
They received more than twenty commentaries.
Two ``out-of-season'' blog posts, one from August 9, 2012 and another from October 22, 2012,
separate first day posts from last posts.
Both have about ten commentaries.
Last blog posts occurred as a few days burst and a final message, a month after.

There are three more recent blog posts, from November and December, 2013.
But these already address \ops\ conception from \vcps.

\subsection{Discussions and etherpads}
Besides blog registries of collective elaborations, four \etherpad s
were written~\cite{etherpads} 
(these are interfaces that allows writing online texts with multiple simultaneous contributors~\cite{etherpads2}):
\begin{itemize}
    \item A pad for important words.
    \item A pad dedicated to a second phase of \vcps\ elaboration, which did not happen yet.
    \item A pad for process documentation. It became the first document described in Section~\ref{refDocs}. 
    \item A pad for both vocabulary specification and ``questions not addressed to in the webinar''.
\end{itemize}
 
\bibliography{plos}
\bibliographystyle{plain}
\end{document}